# Structural, dielectric, and ferroelectric characteristics of the low-temperature sintered 65PMN-35PT sample for electroceramic applications

B. Ramachandran[1,2,*], N. Sudarshan[2,3], G. Mangamma[4], and M.S. Ramachandra Rao[2]

[1]Physics Research Center, Department of Physics, Sethu Institute of Technology, Pulloor, Kariapatti-626115, Tamil Nadu, India.

[2]Department of Physics and Nano Functional Materials Technology Centre, Indian Institute of Technology Madras, Chennai-600036, Tamil Nadu, India

[3]Department of Sustainable Energy Engineering, Indian Institute of Technology Kanpur, Kanpur-208016, Uttar Pradesh, India

[4]Materials Science Group, IGCAR, HBNI, Kalpakkam, Tamil Nadu, India-603102.

[*] Corresponding author: ramskovil@gmail.com (BR)

**ABSTRACT**

A single-phase $0.65PbMg_{1/3}Nb_{2/3}O_3$-$0.35PbTiO_3$ (65PMN-35PT) ceramic was synthesized at a relatively low temperature (875 $^o$C) using a modified columbite method. X-ray diffraction analysis confirmed the single-phase formation of perovskite 65PMN-35PT with a tetragonal structure. Morphological studies indicated that the sample consisted of small grains with a size of ≤ 2 $\mu$m. The dielectric properties of the material demonstrate its relaxor behavior near the ferroelectric transition temperature, $T_C$ = 457 K. The saturation and remnant polarization values of approximately 25.9 and 20.1 $\mu$C cm$^{-2}$ were achieved for an electrically poled sample. Additionally, the poling induced a negative internal electric field of about -0.2 kV cm$^{-1}$ was detected due to the presence of ferroelectric nano-grains in this bulk 65PMN-35PT sample. These observed characteristics of the pyrochlore-free 65PMN-35PT ceramic are similar to those of its single-crystal counterpart.

## 1. Introduction

Lead, Pb-based ferroelectric materials are the most frequently explored and applied materials due to their excellent piezoelectric properties [1]. The most commonly used Pb-based ferroelectrics by researchers and industrialists are Pb(Zr,Ti)$O_3$ (PZT) and PbMg$_{1/3}$Nb$_{2/3}$O$_3$-PbTiO$_3$ (PMN-PT). A material called PbMg$_{1/3}$Nb$_{2/3}$O$_3$ (PMN) is a prototype relaxor ferroelectric with a cubic $A(B′B″)O_3$ perovskite structure, in which the B-sites are occupied by two heterovalent cations [1]. Chemical inhomogeneity is a characteristic structure of relaxor ferroelectrics, leading to a frequency dispersion of the dielectric maximum at a ferroelectric transition temperature near -10 $^o$C [2]. This system is very useful in various device applications, such as multilayer capacitors, actuators, sensors, data storage, and optoelectronic devices [3-5]. Importantly, the PMN-based material exhibits better dielectric, ferroelectric, and piezoelectric properties along with good voltage stability.

Relaxor ferroelectric PMN-based materials with a phase group of $Pm3m$ have a cubic structure over a range of temperatures [1,4]. The addition of PbTiO$_3$ (PT) to PMN forms a binary solution known as (1-$x$)PMN-$x$PT. Consequently, the transition temperature increases with an increase in mole% of PT in this system [5]. The morphotropic phase boundary (MPB) compound materializes at $x$~0.35 [6], separating the rhombohedral (relaxor ferroelectric) phase from the tetragonal (normal ferroelectric) phase of the (1-$x$)PMN-$x$PT system. Overall, this system shows dielectric relaxor behavior for $x$ < 0.3–0.35 and normal ferroelectric behavior for $x$ > 0.35. It displays superior dielectric and piezoelectric characteristics near the MPB. However, a main issue with the solid-state technique for synthesizing PbTiO$_3$-based materials is the high sintering temperature, >1100 $^o$C [7]. Additionally, the high-temperature sintering process results in Pb-loss from the PMN-PT system during synthesis, leading to the formation of pyrochlore impurity phases such as rhombohedral Pb$_2$Nb$_2$O$_7$ or cubic Pb$_3$Nb$_4$O$_{13}$ [8]. These impurity phases are secondary phases of raw materials PbO and Nb$_2$O$_5$.

In recent times, a columbite two-stage calcination method has been utilized [3,9,10], allowing for the synthesis of ferroelectric materials at lower sintering temperatures. This method prevents Pb-loss and the formation of impurity secondary phases. This method involves changing the order of a mixture of the oxide precursors, introducing PbO after MgO to avoid its reaction with $Nb_2O_5$. This prevents the formation of a pyrochlore phase. Therefore, we employed the columbite route to synthesize single-phase and dense 65PMN-35PT ceramics at low temperatures with the help of a sintering aid, LiF. Subsequently, we measured and investigated the dielectric and ferroelectric properties of the 65PMN-35PT ceramic. Despite the low-temperature synthesis, the 65PMN-35PT ceramic exhibited excellent dielectric and ferroelectric characteristics comparable to its single-crystal and thin film counterparts.

## 2. Methodology

Polycrystalline $0.65PbMg_{1/3}Nb_{2/3}O_3$-$0.35PbTiO_3$ (65PMN-35PT) was prepared using a modified two-step columbite precursor method as described in Ref. 10, utilizing the native oxides (as precursors) of Pb, Mg, Nb, and Ti with a purity of ≥99.9%. Initially, the oxides of MgO, $Nb_2O_5$, and $TiO_2$ were finely ground and calcined at 1100 $^o$C for 4 hours. The addition of MgO in this step facilitates its ability to react and mix homogeneously with $Nb_2O_5$, resulting in the formation of a columbite precursor compound, $MgNb_2O_6$. Subsequently, the precursor powder was mixed and ground with stoichiometric PbO content, and the mixture was calcined again at 850 $^o$C for 6 hours to ensure complete diffusion and mixing of PbO with the columbite precursor. Finally, 5 mol% of LiF was added as a flux to the calcined powder and thoroughly mixed using an agate mortar. The final mixture was then sintered in pellet form at low temperatures of $T_s$ = 850–925 $^o$C. This sintering process leads to the material melting locally to promote cohesion and densification of the sample due to the flux-induced high thermal energy during the sintering.

For the sintering process, an aqueous solution (5 at%) of polyvinyl alcohol (PVA) was added to the calcined powder, ground properly, and then pressed into pellets (with a diameter of ~12 mm and thickness of ~1 mm) under a hydrostatic pressure of about ~80 kg/cm$^2$. The sample pellets were pre-heated at 300 $^o$C for 3 hours and then heated at 600 $^o$C for 3 hours to burn out the PVA. Later, the sintering of the pellets was carried out at various temperatures of $T_s$ = 875–925 °C for 1 hour, with a slow heating rate of 5 °C/min. Finally, the sintered sample pellets were cooled down to room temperature using the furnace cooling process.

The pellets obtained from the sample at different sintering temperatures were first characterized using X-ray diffraction (XRD) with the help of a PANalytical X'Pert Pro X-ray diffractometer with Cu $K\alpha$ radiation (wavelength, $\lambda$ = 1.5406 Å). This study was used to obtain a single-phase sample of the perovskite 65PMN-35PT. Afterward, the morphology and elemental characterization of the acquired phase-pure sample were conducted by scanning electron microscopy (SEM) and energy-dispersive X-ray analysis (EDX). The topography response of the sample surface was also studied with atomic force microscopy (AFM) using a scanning probe microscope (SPM, NT-MDT, Russia). For the electrical characterization, the surfaces of the sample pellet were polished and coated with commercial silver (Ag) paste for the top and bottom electrodes. The parallel-plate capacitance ($C_P$) measurement on the sample in the temperature range of 30–300 °C was performed with an applied AC voltage of 0.5 V at different frequencies using an Agilent 4294A impedance analyzer. Finally, the characteristics of the ferroelectric polarization-electric field (*P-E*) and current density-electric field (*J-E*) were measured for the sample at room temperature using a commercial Precision II ferroelectric loop tracer (Radiant Technology, USA).

## 3. Results and Discussion

### 3.1 Structural studies using XRD, SEM, EDX, and AFM

The powder XRD diffractograms of the 65PMN-35PT pellets sintered at different temperatures are shown in Fig. 1. A single-phase perovskite 65PMN-35PT in polycrystalline form with a tetragonal structure (space group *P4mm*) was obtained at a sintering temperature of $T_s$ = 875 °C (Fig. 1a). We attribute the formation of a single-phase at low sintering temperatures to the liquid-phase sintering mechanism during the flux-assisted sintering process [10,11]. However, the secondary pyrochlore phase $Pb_2Nb_2O_7$ begins to form at a slightly higher temperature, $T_s \geq 900$ °C, as indicated by the asterisks in Figs. 1b and 1c. The impurity pyrochlore phase in these high-temperature sintered samples is less than 5% of their parent perovskite phase.

Figure 2 shows the Rietveld refined XRD pattern of the phase-pure 65PMN-35PT sample using the general structure analysis system (GSAS) software [12]. A good fit is achieved with fitting factors of $\chi^2$ = 3.238, $wRp$ = 8.1%, and $Rp$ = 6%. The difference profile between the calculated and observed diffraction data is shown at the bottom of Fig. 2. The deduced lattice parameters of the sample are $a$ = 4.012(1) Å and $c$ = 4.053(1) Å with a tetragonality factor of $c/a$ = 1.01. The density of the synthesized sample pellets was determined using the Archimedes method. The sintering density of the sample prepared at $T_s$ = 875 °C is approximately 7.63 g cm$^{-3}$. This value is slightly higher than the other two samples (≤7.12 g cm$^{-3}$) sintered at slightly higher temperatures $T_s$ ≥900 °C (see inset of Fig. 1b). This finding indicates that the sample sintered at low-temperature $T_s$ = 875 °C is highly dense (with a relative density of ~94%) compared to the other two samples (≤88%). The relative density was estimated by comparing the measured density value to its theoretical density (8.08 g cm$^{-3}$) obtained from the Rietveld refinement of the XRD data. Using the Williamson-Hall relation [13], $\beta cos\theta = \frac{K\lambda}{D} + 4\varepsilon sin\theta$, we estimated the crystallite size ($D$) and lattice strain ($\varepsilon$) values for the bulk 65PMN-35PT sample to be approximately 42 nm and 0.1%, respectively. Here, $\beta$ represents the full-width at half maximum (FWHM) of the diffraction peak, and $K$ = 0.9 is the shape factor.

The inset of Fig. 2 presents the SEM image of the polished ceramic sample after HF etching. The dense formation of small grains is observed in the 65PMN-35PT ceramic sample with an average grain size of ~1.3 $\mu$m. Since the present sintering process is rapid, i.e., completed in only one hour, it results in microstructural inhomogeneity and very small grains compared to the conventional sintering process [14]. The composition of the sample was computed from the EDX data. The deduced atomic percentages are about 19.6, 4.3, 9.2, 8.1, and 58.8% for the elements Pb, Mg, Nb, Ti, and O. We found that the measured composition of these elements in the 65PMN-35PT sample is close to their nominal compositions of 20.0, 4.3, 8.7, 7.0, and 60.0%, respectively.

Figure 3 showcases the 2-dimensional (2D) AFM topography image of the investigated 65PMN-35PT pellet. The AFM study verifies that the ceramic sample consists of small grains with sizes ranging from 1.0 to 2.0 $\mu$m, consistent with the SEM analysis (refer to the inset of Fig. 2). The estimated average surface roughness of the 65PMN-35PT ceramic is approximately 231 nm. For this AFM study, a Pt-coated Si tip with stiffness, 16 N m$^{-1}$ and a diameter of 50 nm was used for topography imaging, with a scanning speed of 1.1 Hz. The AFM study reveals micron-sized void defects in the bulk 65PMN-35PT sample, likely due to its sintering density being only about 94%. Therefore, the remaining 6% of space in the sample is occupied by void defects as detected in the AFM study. However, no cracks were observed across the entire surface area of the sample. Our unpublished piezo force microscopy study revealed that the switching of ferroelectric domains was confirmed from the contrast PFM images (from dark to bright) by changing the applied DC bias voltage from +10 V to -10 V (the figure is not shown here). Interestingly, a ferroelectric domain structure with a 109$^o$ domain wall type was witnessed for the 65PMN-35PT sample, which is consistent with the ferroelectric domain structures of 70PMN-30PT single crystals with different domain wall types (180$^o$/71$^o$, 180$^o$/109$^o$, and 109$^o$) reported in the literature [15]. The detailed PFM analysis, along with an atomic force acoustic microscopy (AFAM) study on the inspected 65PMN-35PT sample, will be conducted and published elsewhere.

**3.2 Temperature- and frequency-dependent dielectric properties**

The semi-log plots of the frequency (*f*)-dependent dielectric constant ($\varepsilon_r$) and dielectric loss (*tanδ*) of the 65PMN-35PT ceramic measured at room temperature are displayed in Fig. 4a. It was observed that the dielectric constant ($\varepsilon_r$) decreases gradually with increasing frequency, while the dielectric loss (*tanδ*) increases with frequency up to 13.8 kHz, after which it decreases steadily with frequency. Specifically, the dielectric loss (*tanδ*) of the sample at sample frequencies of 1 kHz, 11 kHz, and 70 kHz was found to be approximately 0.01, 0.03, and 0.02, respectively. The corresponding dielectric constants at these frequencies are approximately 32.9 × 10$^3$, 1.5 × 10$^3$, and 0.3 × 10$^3$, respectively. To analyze the different contributions, such as grains and grain boundaries, to the complex dielectric data ($\varepsilon^* = \varepsilon'_r + i\varepsilon''_r$) of the 65PMN-35PT compound, a Cole-Cole plot between the real dielectric constant ($\varepsilon'_r$) and imaginary dielectric constant ($\varepsilon''_r$) at room temperature was created and is presented in Fig. 4b. A spike feature was witnessed at low frequencies up to 2 kHz, likely due to electrode-related polarization processes [16]. In the intermediate frequency region, an observed semicircle feature centered near 11 kHz is attributed to the contribution of grain boundaries in the sample. Finally, the contribution from the polarization of grains to the sample is observed at high frequencies, $f > 1$ MHz (see inset of Fig. 4b).

Figure 5a illustrates the data of the dielectric constant $\varepsilon_r(T)$ versus temperature for the inspected MPB compound 65PMN-35PT at different frequencies. The sample shows a peak feature in the $\varepsilon_r(T)$ data near a ferroelectric-paraelectric transition temperature, $T_C$ = 184 °C (457 K) at 11 kHz, similar to the literature data ($T_C$ = 454 K) [17]. A fairly narrow frequency dispersion near the dielectric maxima (i.e., near $T_C$) indicates that our sample has an ordered perovskite structure [18]. Additionally, its $\varepsilon_{r\text{-max}}$ value is ~5900 near $T_C$, which is lower compared to the large value ($\geq 10^4$) of some PMN-PT-based bulk materials reported in the literature [6,16-18]. However, its value is in the same order as that of its heteroepitaxial film ($\varepsilon_{r\text{-max}}$~4800) and the (110) and (111)-oriented single crystal ($\varepsilon_{r\text{-max}}$~4500) counterparts [20,21]. These findings can be explained by a *rattling ion model*, which states that large *B*-site ions (i.e., Nb-ions) cannot move easily because of the small rattling space available in the ordered perovskite under an applied electric field. Thus, the ordered system has a smaller $\varepsilon_r$ [18], which is the case for our sample. The deduced values of the parameters $T_C$ and $\varepsilon_{r\text{-max}}$ of the studied 65PMN-35PT sample, along with the literature values, are listed in Table 1.

However, the *B*-site ions (particularly high-valence Nb-ions) can move easily without distorting the octahedral because of their large rattling space in the disordered perovskite systems [16-18], and hence they have a larger $\varepsilon_r$ and Curie-Weiss constant [18]. Furthermore, we do not observe the low-*T* monoclinic-tetragonal phase transition below $T_C$, which was previously observed in this MPB compound, 65PMN-35PT [21-23]. These findings confirm the high-quality nature of our 65PMN-35PT sample with a tetragonal-ordered perovskite structure, as evidenced by our structural and dielectric studies (see Figs. 1-4).

To check whether the present compound is undergoing dielectric relaxation or not, we employed a modified Curie-Weiss law [18,22], which is described by the relation

$$\frac{1}{\varepsilon_r} = \frac{1}{\varepsilon_{r-max}} + \frac{(T-T_C)^2}{2\varepsilon_{r-max}\delta^2}. \tag{1}$$

Where $\varepsilon_{r-max}$ and $\delta$ parameters correspond to the maximum dielectric constant near $T_C$ and the diffuseness parameter. A plot of $1/\varepsilon_r$ versus $(T-T_C)^2$ at two representative frequencies of 20 and 70 kHz is presented in Fig. 5b, and the solid lines are the corresponding linear fits. The resultant linear fit signifies that the sample has dielectric relaxor behavior with a diffusive-type ferroelectric transition. We estimated the diffuseness parameter ($\delta$) to be about 51 K at 20 kHz, which was found to vary weakly (48–52 K, Table 1) with increasing frequency. Moreover, the Curie temperature increases gradually (184–194 °C) with an increase in frequency. While the $\varepsilon_{r-max}$ value decreases slightly with increasing frequency, from ~5900 to ~5222.

**3.3 Ferroelectric PE loop and leakage current studies**

The ferroelectric hysteresis loops, displaying polarization ($P$) versus electric field ($E$), of unpoled 65PMN-35PT pellets sintered at different temperatures were measured at an applied voltage of 4 kV and are shown in Fig. 6a. We estimated the values of saturation polarization ($P_s$), remnant polarization ($P_r$), coercive electric field ($E_c$), and internal electric field ($E_I$) for these samples, which are listed in Table 1. It is observed that the saturation polarization ($P_s$ = 22.3 $\mu$C cm$^{-2}$) and remnant polarization ($P_r$ =14.0 $\mu$C cm$^{-2}$) of the 65PMN-35PT sample sintered at $T_s$ = 875 °C are relatively higher compared to values ($P_s$ ≤20.0 and $P_r$ ≤12.5 $\mu$C cm$^{-2}$) of samples sintered at high temperatures $T_s$ ≥ 900 °C. Similarly, the coercive field ($E_c$ = 11.3 kV cm$^{-1}$) of the low-temperature prepared sample is slightly higher compared to samples sintered at high temperatures. However, the deduced internal field ($E_I$ = 0.25 kV/cm) of samples sintered at $T_s$ = 875 °C is lower compared to the other two samples ($E_I$ ≥0.45 kV/cm, see Table 1).

The ferroelectric $P$-$E$ characteristics of the unpoled and poled 65PMN-35PT pellet samples prepared at $T_s$ = 875 °C recorded at an applied voltage of 4 kV with a testing frequency of ~40 Hz are presented in Fig. 6b for comparative analysis. The electrical poling of the sample was conducted with an applied electric field of $E$ = 20 kV cm$^{-1}$ at 80 °C for 1 h. For the unpoled sample, the saturation polarization ($P_s$) and remnant polarization ($P_r$) are about ~22.3 and ~14.0 $\mu$C cm$^{-2}$, respectively. After electrical poling, both $P_s$ and $P_r$ increased to values of ~25.9 and ~20.1 $\mu$C cm$^{-2}$, respectively. Interestingly, the $P_s$ and $P_r$ values of our ceramic sample are slightly larger than the bulk 65PMN-35PT sample prepared by the co-precipitation method (see Table 1) [19]. These values are also similar to those of the heteroepitaxial 65PMN-35PT thin film [21]. These observations are likely due to the poling-induced alignment of electric dipoles and an enhancement in ferroelectric domain mobility in the inspected material. Thus, improved ferroelectric properties were observed for the 65PMN-35PT material via electric poling [24,25]. Additionally, the $P_s$ and $P_r$ values of this sample are slightly higher than samples sintered at higher temperatures, $T_s \geq 900$ °C (see Fig. 6a). Furthermore, our estimation reveals a higher energy density loss ($E_{loss}$ = 0.33 J cm$^{-3}$) for the poled sample compared to the unpoled sample (~0.25 J cm$^{-3}$). However, the $E_{loss}$ value of the unpoled sample is more than two times higher when compared with a soft ferroelectric 90PMN-10PT ceramic [26]. These findings suggest that our 65PMN-35PT sample is a hard ferroelectric material.

An asymmetry was observed in the *P-E* curve of both samples. The unpoled sample shows a slight shift in the positive electric field direction, while the poled sample shift in the negative field direction. This finding is attributed to the occurrence of an internal electric field in this piezoelectric ceramic [26-28]. The positive coercive fields (+$E_C$) obtained for the unpoled and poled samples are about 11.3 and 12.9 kV cm$^{-1}$, respectively. We also calculated the internal electric fields ($E_I$) of these samples to be between +0.25 and -0.20 kV cm$^{-1}$, respectively. Interestingly, the observed negative internal field ($E_I$) for the poled sample is likely due to interactions between polar nano-grains (see Fig. 3) and/or built-in charge resulting from disorders that lead to the inhomogeneity in ferroelectric domain structures [29]. However, these internal field values are significantly lower when compared to PMN-PT-based and acceptor Mn-doped PZT piezoelectric materials (>1.0 kV cm$^{-1}$; see Table 1) [25-27]. We recognize that the internal electric field $E_I$ of piezoelectric systems is directly related to their electromechanical quality factor, $Q_m$ [28]. Therefore, the presence of the internal field $E_I$ could potentially enhance the $Q_m$ of the piezoelectric materials. Consequently, we propose that our 65PMN-35PT ceramic sample could be utilized in various piezoceramic devices, especially in high-power acoustic applications.

Additionally, a comparison of the *P-E* loop of columbite-synthesized 65PMN-35PT with the solid-state synthesized sample is illustrated in Fig. 6c. It is noted that the sample prepared by the two-step columbite method has a better hysteresis loop compared to the solid-state prepared sample. Importantly, the $P_s$ and $P_r$ values ($P_s$ = 22.3 μC cm$^{-2}$ and $P_r$ = 14.0 μC cm$^{-2}$) of the columbite-synthesized sample are higher than those of the solid-state-synthesized sample ($P_s$ = 18.2 μC cm$^{-2}$ and $P_r$ = 11.7 μC cm$^{-2}$). The reported sample has a slightly lower coercive field ($E_C$ = 11.3 kV cm$^{-1}$) compared to the solid-state-prepared sample ($E_C$ = 13.9 kV cm$^{-1}$). Under the same applied electric field, the columbite-synthesized sample exhibits better polarizability than the sample prepared by the solid-state route. These characteristics of the columbite-prepared sample are attributed to its single-phase ordered perovskite structure, which has better dielectric and ferroelectric characteristics. Hence, this ceramic sample can be utilized for electroceramic device applications. In contrast, the sample prepared by the solid-state route has a secondary pyrochlore phase (Pb$_2$Nb$_2$O$_7$) of about ~11% (figure not shown here) and is slightly more difficult to polarize electrically compared to the columbite-synthesized sample.

To analyze the leakage current characteristics of the 65PMN-35PT ceramics, we recorded the current density (*J*) versus the electric field (*E*), as shown in Fig. 7. The sample displays butterfly-like *J-E* characteristics, typical of good dielectric material. The sample exhibits a low leakage current density of 1.6 × 10$^{-5}$ A cm$^{-2}$ at a maximum applied field of 40 kV cm$^{-1}$. The sample demonstrates ohmic conduction with a power factor (*n*) of about 1.4 at low fields < 11 kV cm$^{-1}$ (the inset of Fig. 7). With further increasing field, the sample exhibits space-charge-limited conduction, as described by Mott's equation [30,31]:

$$J = \frac{9}{8}\varepsilon_0\varepsilon_{rs}\theta\mu\frac{V^n}{d^3}. \qquad (2)$$

Here, the factors $\varepsilon_0$, $\varepsilon_{rs}$, and $\mu$ represent the permittivity of free space, the static dielectric constant, and the mobility of the electrons in the sample. The parameter $\theta$ denotes the ratio of the number of free electrons to the number of trapped electrons. The parameters *d* and *V* correspond to the distance between electrodes (i.e., the sample thickness) and the applied voltage.

The inset of Fig. 7 illustrates the log-log *J-E* plot, and the solid lines correspond to the linear fits. At low electric fields of <20 kV cm$^{-1}$, the sample exhibits ohmic conduction with *n* = 1.4. Notably, the sample shows space-charge-limited conduction with a power factor of *n* = 2.0 at higher electric fields >25 kV cm$^{-1}$. This finding indicates a steady-state conduction mechanism for the sample at higher fields. In the intermediate fields, a negative inflection point in the *J-E* data is observed near *E* = 20 kV cm$^{-1}$. This is likely due to the electrons being trapped on the sample surfaces because of the volatilization of the sintering aid LiF, which leads to the observed negative electrical resistivity for the sample at particular applied fields. Further study is needed to clarify this observation.

In brief, we have synthesized a single-phase and highly dense 65PMN-35PT ceramic sample at a lower sintering temperature. Our sample exhibits excellent ferroelectric and dielectric characteristics with a low leakage current density and a small internal bias electric field, similar to the PMN-PT-based crystals and epitaxial film [21,22]. Therefore, the synthesized single-phase ferroelectric bulk 65PMN-35PT material could be a valuable material for practical electroceramic applications [32-34].

## 4. Conclusion

A single-phase, highly dense 65PMN-35PT pellet sample was obtained at a low sintering temperature (875 °C) through the two-step columbite precursor route. Our X-ray diffraction analysis showed that the synthesized sample crystallized in a tetragonal structure (*P4mm*) with a tetragonality factor of $c/a = 1.01$. Importantly, the sample consists of small grains with microstructural inhomogeneity due to its lower sintering temperature and short sintering period. From the dielectric and ferroelectric data, we observed that this material exhibits very good dielectric and ferroelectric properties. Notably, the electrically poled sample demonstrates excellent ferroelectric characteristics, with saturation and remnant polarization values of $P_s = 25.9$ and $P_r = 20.1$ μC cm$^{-2}$, respectively, compared to the unpoled sample ($P_s = 22.3$ and $P_r = 14.0$ μC cm$^{-2}$) and the sample prepared by the solid-state reaction ($P_s = 18.2$ μC cm$^{-2}$ and $P_r = 11.7$ μC cm$^{-2}$). Additionally, the unpoled sample exhibits a small internal electric field of $E_I = +0.25$ kV cm$^{-1}$ and a coercive field of $E_C = 11.3$ kV cm$^{-1}$. Interestingly, the poled 65PMN-35PT sample shows a negative internal electric field of $E_I = -0.20$ kV cm$^{-1}$ and a higher coercive field of $E_C = 12.9$ kV cm$^{-1}$, attributed to the nano ferroelectric grains in the sample. Furthermore, the studied sample demonstrates a low leakage current density (*J*) of approximately $1.6 \times 10^{-5}$ A cm$^{-2}$ at an applied field of $E = 40$ kV cm$^{-1}$. These observed characteristics suggest that the sample is a tetragonally ordered ferroelectric material with excellent dielectric and ferroelectric properties. Therefore, this investigated hard ferroelectric 65PMN-35PT ceramic could prove to be very useful in electroceramic applications.

**Credit Authorship contribution statement**

**B. Ramachandran:** Writing–Original Draft, Writing–Review & Editing, Formal Analysis, Data Curation, Conceptualization. **N. Sudarshan:** Writing–Review & Editing, Formal Analysis, Data Curation. **G. Mangamma:** Writing–Review & Editing, Formal Analysis, Data Curation. **M.S. Ramachandra Rao:** Writing–Review & Editing, Project Administration, Resources.

**Funding:** This research work did not receive any financial support from funding agencies in the public, commercial, or not-for-profit sectors.

**Data Access Statement:** Data are available on request from the authors.

**ORCID ID:**

B. Ramachandran - https://orcid.org/0000-0003-3676-1410

## Declarations

## References

1. R.A. Cowley, S.N. Gvasaliya, S.G. Lushnikov, *et al.*, Relaxing with relaxors: a review of relaxor ferroelectrics, Adv. Phys. **60**, 229-327 (2009). doi: 10.1080/00018732.2011.555385
2. S. Fengbing, L. Qiang, Z. Haisheng, *et al.*, Phase formation and transitions in the lead magnesium niobate-lead titanate system, Mat. Chem. Phys. **83**, 135–139 (2004). doi: 10.1016/j.cap.2005.11.006
3. R.F. Mamin, D. A. Tayurskii, Phenomenological model of dielectric properties of PMN-PT, Ferroelectrics **509**, 27-31 (2017). doi: 10.1080/00150193.2017.1292110
4. Y. Huang, L. Zhang, W. Shi, et al., Ferroelectric-to-relaxor transition and ultra-high electrostrictive effect in $Sm^{3+}$-doped $Pb(Mb_{1/3}Nb_{2/3})O_3$-$PbTiO_3$ ferroelectrics ceramics, J. Mater. Sci. Tech. **165**, 75-84 (2023). doi: 10.1016/j.jmst.2023.04.046
5. S.I. Shkuratov, C.S. Lynch, A review of ferroelectric materials for high power devices, J. Materiomics **8**, 739-752 (2022). doi: 10.1016/j.jmat.2022.04.002
6. E.V. Colla, N.K. Yushin, and D. Viehland, Dielectric properties of $(PMN)_{(1-x)}(PT)_x$ single crystals for various electrical and thermal histories, J. Appl. Phys. **83**, 3298–3304 (1998). doi: 10.1063/1.367098
7. S. Kustov, J.M. Obrero, X. Wang, D. Damjanovic, E.K.H. Salje, Phase transitions in the ferroelectric relaxor $(1-x)Pb(Mb_{1/3}Nb_{2/3})O_3$-$xPbTiO_3$, Physical Review Materials **6**, 124414 (2022). doi: 10.1103/PhysRevMaterials.6.124414
8. E.R. Camargo, M. Kakihana, E. Longo, *et al.*, Pyrochlore-free $Pb(Mg_{1/3}Nb_{2/3})O_3$ prepared by a combination of the partial oxalate and the polymerized complex

methods, J. Alloys Compounds **314**, 140–146 (2001). doi: 10.1016/S0925-8388(00)01220-2

9. J. Yoo, I.H. Lee, D.S. Paik, *et al.*, Piezoelectric and dielectric properties of low temperature sintering $Pb(Mn_{1/3}Nb_{2/3})O_3$-$Pb(Zn_{1/3}Nb_{2/3})O_3$-$Pb(Zr_{0.48}Ti_{0.52})O_3$ ceramics with variation of sintering time, J. Electroceram. **23**, 519–523 (2009). doi: 10.1007/s10832-008-9524-0

10. B. Ramachandran, N. Sudarshan, M.S. Ramachandra Rao, Magnetoimpedance and magnetodielectric properties of single phase 45PMN-20PFW-35PT ceramics, J. Appl. Phys. **107**, 09C503 (2010). doi: 10.1063/1.3355543

11. X.Y. Tong, J.J. Zhou, H. Liu, *et al.*, The electrical properties of low-temperature sintered PMN-PT electrostrictive ceramics by LiF modification, J. Mater. Sci.: Mater. Electron. **27**, 10729–10734 (2016). doi: 10.1007/s10854-016-5174-1

12. A.C. Larson and R.B. Von Dreele, Los Alamos National Laboratory, Report No. LAUR. 86-748, 1994.

13. B. Ramachandran, A. Dixit, M.S. Ramachandra Rao, Investigation of electronic energy levels in a weak ferromagnetic oxygen-deficient $BiFeO_{2.85}$ thick film using absorption and X-ray photoelectron spectroscopic studies, Surface and Interface Analysis (2025). doi: 10.1002/sia.7399

14. R. Zuo, T. Granzow, D.C. Lupascu, *et al.*, PMN-PT ceramics prepared by spark plasma sintering, J. Am. Ceram. Soc. **90**, 1101–1106 (2007). doi: 10.1111/j.1551-2916.2007.01533.x

15. D. Zhang, L. Wang, L. Li, P. Sharma, J. Seidel, Varied domain structures in $0.7Pb(Mg_{1/3}Nb_{2/3})O_3$-$0.3PbTiO_3$ single crystals, Microstructures **3**, 2023046 (2023). doi: 10.20517/microstructures.2023.57

16. B. Behera, E.B. Araujo, R.N. Reis, J.D.S. Guerra, AC conductivity and impedance properties of $0.65Pb(Mg_{1/3}Nb_{2/3})O_3$-$0.35PbTiO_3$ ceramics, Advances in Condensed Matter Physics **2009**, 1-6 (2009). doi: 10.1155/2009/361080

17. H Wang, Z He, X. Wang, *et al.*, Preparation and characterization of pyrochlore-free $0.655Pb(Mg_{1/2}Nb_{2/3})O_3$-$0.345PbTiO_3$ piezoelectric ceramics by tap-casting process, Ferroelectrics **531**, 175-185 (2018). doi: 10.1080/00150193.2018.1497421

18. K. Uchino (Edt.), Relaxor ferroelectric-based ceramics, in Advanced Piezoelectric Materials: Science and Technology (Woodhead Publishing Ltd., 2010), pp.116–119. doi: 10.1533/9781845699758.1.111

19. A.N. Tarale, S. Premkumar, V.R. Reddy, *et al.*, Dielectric and ferroelectric properties of $(1-x)Pb(Mg_{1/2}Nb_{2/3})O_3$-$(x)PbTiO_3$ synthesized by co-precipitation method, Ferroelectrics **516**, 8–17 (2017). doi: 10.1016/S0167-577X(01)00527-4

20. M. Allaverdi, A. Hall, R. Brennan, *et al.*, An overview of rapidly prototyped piezoelectric actuators and grain-oriented ceramics, Journal of Electroceramics **8**, 129-137 (2022). doi: 10.1023/A:1020503929340

21. K. Wasa, I. Kanno, T. Suzuki, *et al.*, Structural and ferroelectric properties of single crystal PMNT thin films, Integrated Ferroelectrics **70**, 131–140 (2005). doi: 10.1080/10584580590898767

22. F. Li, S. Zhang, Z. Xu, *et al.*, Composition and phase dependence of the intrinsic and extrinsic piezoelectric activity of domain engineered $(1-x)Pb(Mg_{1/2}Nb_{2/3})O_3$-$xPbTiO_3$ crystals, J. Appl. Phys. **108**, 034106 (2010). doi: 10.1063/1.3466978

23. M.V. Takarkhede and S.A. Band, Synthesis, structural and dielectric properties of 0.8PMN-0.2PT relaxor ferroelectric ceramic, Bull. Mater. Sci. **40**, 917–923 (2017). doi: 10.1007/s12034-017-1444-7
24. C. Qiu, J. Liu, F. Li, Z. Xu, Thickness dependence of dielectric and piezoelectric properties for alternating current electric-field-poled relaxor-PbTiO3 crystals, Journal of Applied Physics **125**, 014102 (2019). doi: 10.1063/1.5063682
25. J.W. Sun, T.T. Zate, W.J. Choi, G.J. Lee, Y.S. Jeong, S.G. Lee, J.E. Ryu, W. Jo, Suppressing phase transformation-induced depolarization in PMN-PT single crystals through high-temperature AC poling, Scripta Materialia **258**, 116519 (2025). doi: 10.1016/j.scriptamat.2024.116519
26. T.F. Zhang, X.G. Tang, Q.X. Liu, *et al.*, Energy-storage properties and high-temperature dielectric relaxation behaviors of relaxor ferroelectric $Pb(Mg_{1/2}Nb_{2/3})O_3$-$PbTiO_3$ ceramics, J. Phys. D: Appl. Phys. **49**, 095302 (2016). doi: 10.1088/0022-3727/49/9/095302
27. H.T. Oh, H.J. Joo, M.C. Kim, *et al.*, Effect of Mn on dielectric and piezoelectric properties of 71PMN-29PT ($71Pb(Mg_{1/2}Nb_{2/3})O_3$-$29PbTiO_3$) single crystals and polycrystalline ceramics, J. Korean Ceram. Soc. **55**, 166–173 (2018). doi: 10.4191/kcers.2018.55.2.04
28. Y. Gao, K. Uchino, D. Viehland, Time dependence of the mechanical quality factor in hard lead zirconate titanate ceramics: Development of an internal dipolar field and high power origin, Jpn. J. Appl. Phys. **45**, 9119–9124 (2006). doi: 10.1143/JJAP.45.9119
29. F. Wu, W. Cai, Y.W. Yeh, S. Xu, N. Yao, Energy scavenging based on a single-crystal PMN-PT nanobelt, Scientific Reports **6**, 22513 (2016). doi: 10.1038/srep22513
30. B. Ramachandran, A. Dixit, R. Naik, *et al.*, Charge transfer and electronic transitions in polycrystalline $BiFeO_3$, Phys. Rev. B **82**, 012102 (2010). doi: 10.1103/PhysRevB.82.012102
31. S.T. Chang, J.Y.M. Lee, Electrical conduction mechanism in high-dielectric-constant $Ba_{0.5}Sr_{0.5}TiO_3$ thin films, Appl. Phys. Lett. **80**, 655 (2002). doi: 10.1063/1.1436527
32. Y. Yan, Z. Li, L. Jin, H. Du, M. Zhang, D. Zhang, Y. Hao, Extremely high piezoelectric properties in Pb-based ceramics through integrating phase boundary and defect engineering, ACS Applied Materials and Interfaces **13**, 38517-38525 (2021). doi: 10.1021/acsami.1c10298
33. H. Ursic, M.S. Zarnik, M. Kosec, $Pb(Mg_{1/2}Nb_{2/3})O_3$-$PbTiO_3$ (PMN-PT) material for actuator applications, Smart Materials Research **2011**, 1-6 (2011). doi: 10.1155/2011/45290
34. A. Saxena, A. Hussain, A. Saxena, A.J. Joseph, R.S. Saxena, Dielectric dispersion near the morphotropic phase boundary of 0.64PMN-0.36PT ceramics, Ceramic Internationals **48**, 26258-26263 (2022). doi: 10.1016/j.ceramint.2022.05.307

**Table 1** The obtained parameters of $T_C$, $\varepsilon_r$, $\varepsilon_{rmax}$, $tan\delta$, $\delta$, $P_s$, $P_r$, $E_{loss}$, $E_C$, and $E_I$ for the poled 65PMN-35PT pellet sintered at 875 °C from the dielectric and ferroelectric measurements, along with the data of unpoled 65PMN-35PT samples and related literature data.

| Parameters | Poled 65PMN-35PT pellet sintered at 875 °C (this work) | Unpoled 65PMN-35PT pellet sintered at 875 °C (this work) | Unpoled 65PMN-35PT pellet sintered at 900 °C (this work) | Unpoled 65PMN-35PT pellet sintered at 925 °C (this work) | Bulk 65PMN-35PT (the literature) | 65PMN-35PT epitaxial film [18] |
|---|---|---|---|---|---|---|
| $T_C$ (K) | 457 | - | - | - | 454 [16] & 438 [20] | 473 |
| $\varepsilon_r$ at 300 K | 1474 | - | - | - | 2720 [17] | 2700 |
| $tan\delta$ at 300 K | 0.03 | - | - | - | - | - |
| $\varepsilon_{rmax}$ near $T_C$ | 5900 | - | - | - | 24698 [17] | 4800 for epitaxial PMN-PT film [18] & 4500 for (110)-oriented PMN-PT crystal [19] |
| Diffuseness factor, $\delta$ (K) | 48-52 | - | - | - | 38-50 [22] | - |
| $P_s$ ($\mu C/cm^2$) | 25.9 | 22.3 | 16.7 | 20.0 | 24.0 [17] | 38.0 |
| $P_r$ ($\mu C/cm^2$) | 20.1 | 14.0 | 12.5 | 10.7 | 18.5 [17] | 20.0 |
| Energy density loss, $E_{loss}$ (J/cm$^3$) | 0.33 | 0.25 | 0.19 | 0.19 | 0.13 [23] (soft ferroelectrics) | - |
| Coercive electric field, $E_C$ (kV/cm) | 12.9 | 11.3 | 11.1 | 9.6 | 10.7 [17] | 50-90 |
| Internal electric field, $E_I$ (kV/cm) | -0.20 | +0.25 | +0.60 | +0.45 | 2.0 (1.2 kV/cm for the crystal [24]) | - |

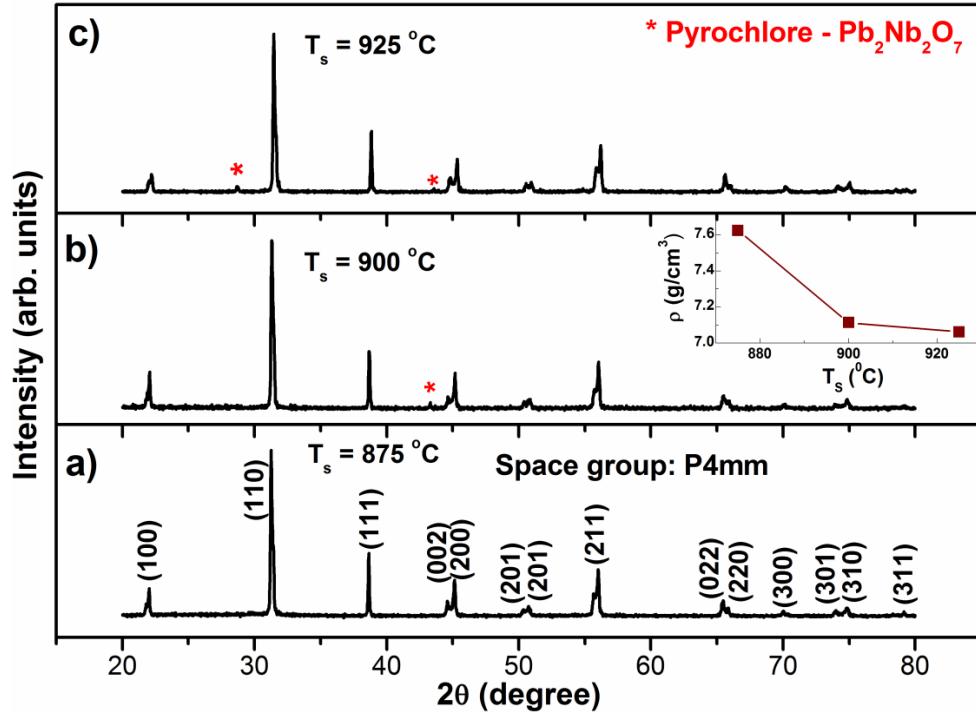

**Fig. 1** The XRD patterns of the prepared 65PMN-35PT samples sintered at different temperatures. The observed pyrochlore phase $Pb_2Nb_2O_7$ is marked by asterisks in Fig. 1b and 1c. The inset of Fig. 1b displays the measured density ($\rho$) of the synthesized samples versus their sintering temperature ($T_s$).

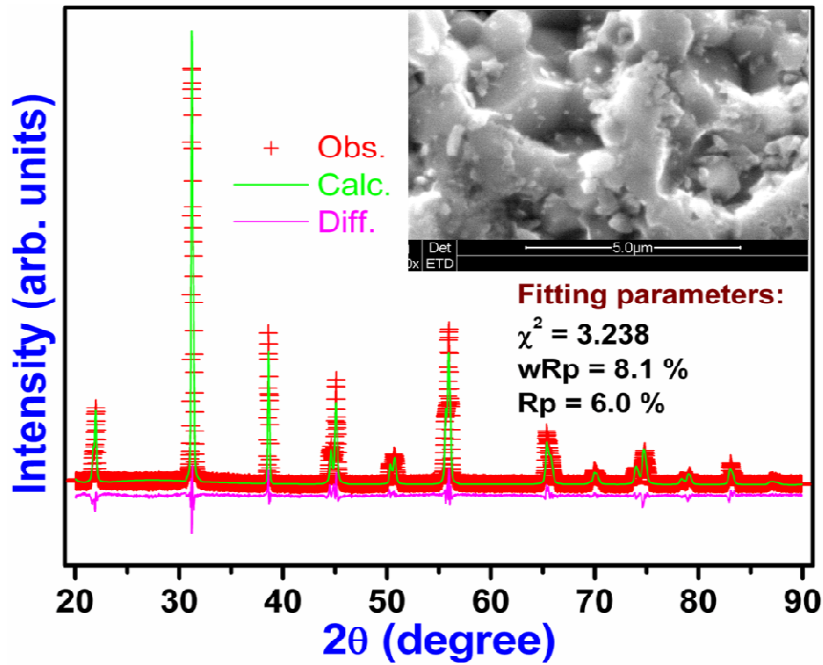

**Fig. 2** The Rietveld refined XRD pattern of the polycrystalline 65PMN-35PT sample sintered at 875 °C. The difference (Diff.) between the observed (Obs.) and calculated (Calc.) patterns is displayed at the bottom. The inset shows the SEM image of the polished pellet.

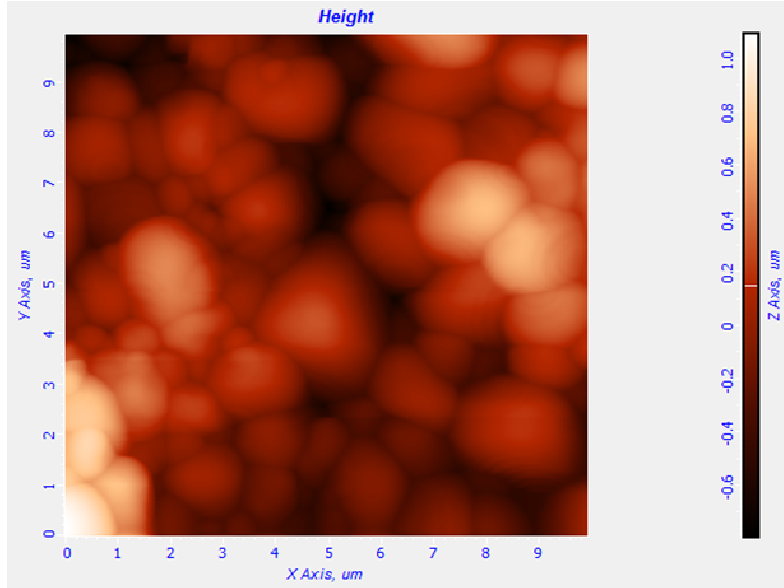

**Fig. 3** The AFM image of the 65PMN-35PT ceramic sample.

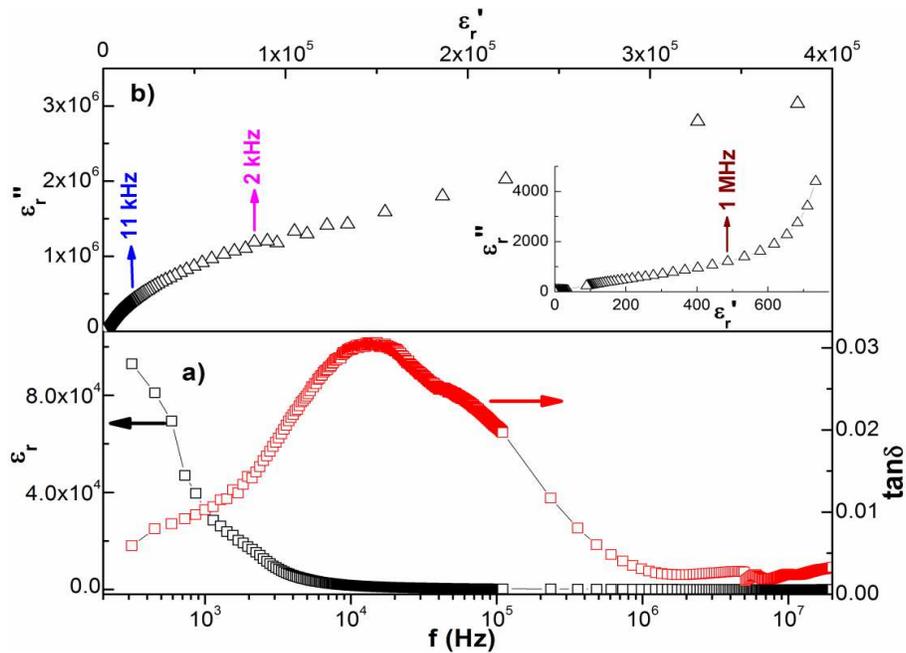

**Fig. 4 a** The semi-log plots of the frequency-dependent dielectric constant ($\varepsilon_r$) and dielectric loss (*tanδ*) of the 65PMN-35PT sample and **b** the Cole-Cole plot using the frequency-dependent real and imaginary dielectric constants ($\varepsilon'_r$ and $\varepsilon''_r$). The inset of Fig. 4b displays the complex dielectric permittivity spectrum in the high-frequency range above 1 MHz.

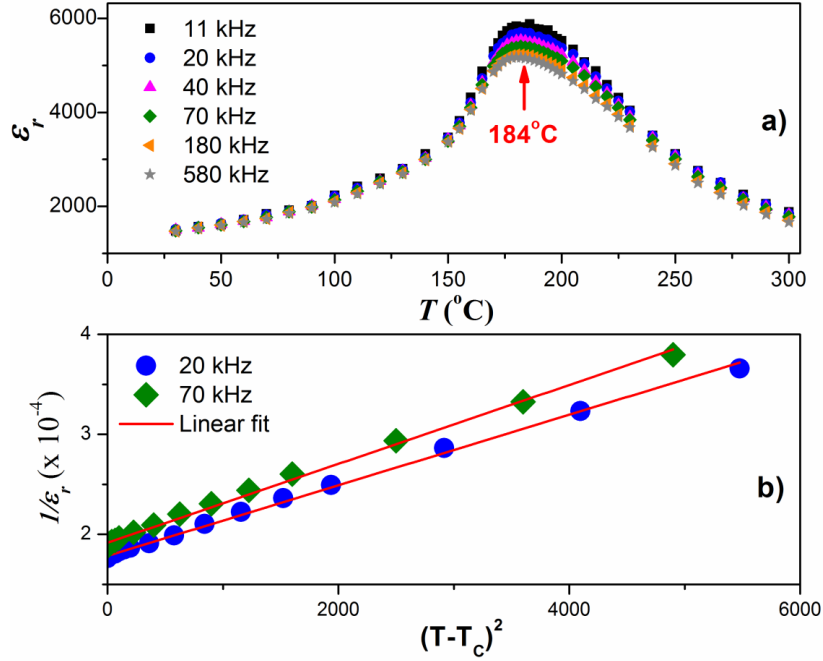

**Fig. 5 a** Temperature-dependent dielectric constant of 65PMN-35PT ceramics at different frequencies and **b** the graph of $1/\varepsilon_r$ versus $(T-T_C)^2$ for the frequencies, 20 and 70 kHz. The straight lines in Fig. 3b correspond to the linear fits using Eq. 1.

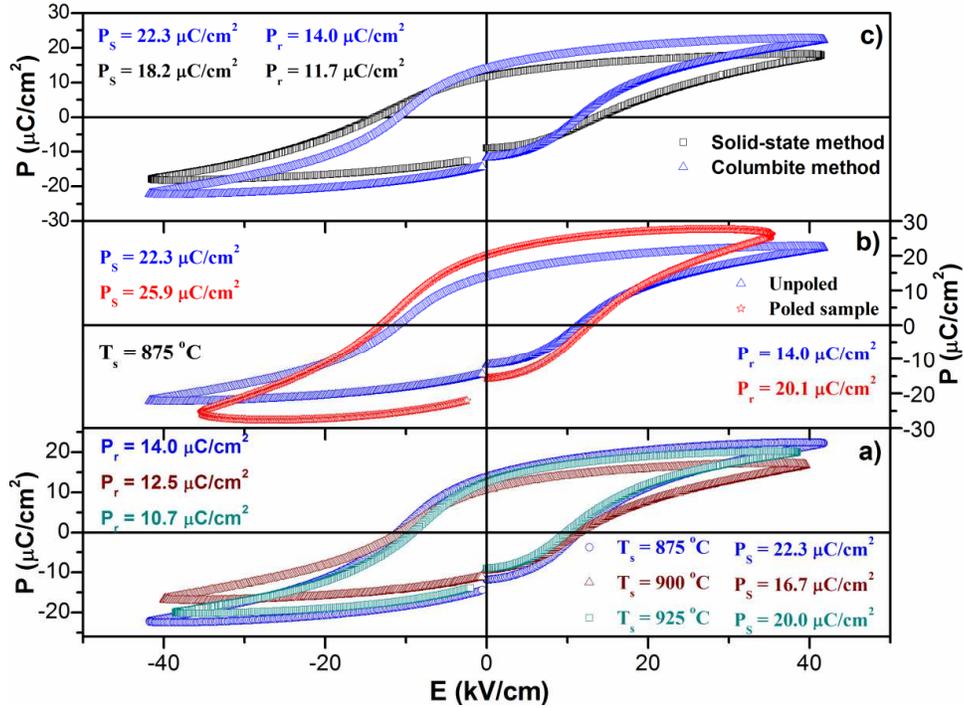

**Fig. 6 a** Ferroelectric *P-E* hysteresis curves of the unpoled 65PMN-35PT pellets sintered at different temperatures, **b** *P-E* hysteresis loops of the unpoled and poled samples synthesized by the columbite method, and **c** the comparison of the columbite-made unpoled 65PMN-35PT ceramic with the unpoled sample prepared by solid-state reaction.

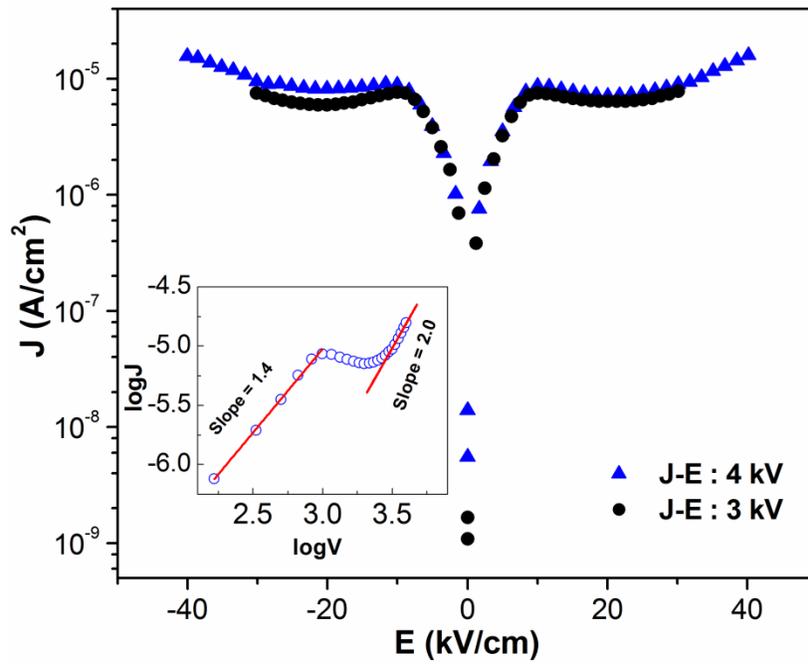

**Fig. 7** *J-E* characteristics for the studied sample. The inset shows the log-log plot of *J* versus *V*, and the straight lines are linear fits to the data using Eq. 2.